\documentclass[sigconf]{acmart}
\usepackage{booktabs} % For formal tables
\usepackage{multirow}
\usepackage{subcaption}
\usepackage{caption}
\usepackage{setspace}
\usepackage{amsmath}
\usepackage[english]{babel}
\usepackage{mathrsfs}
\definecolor{mygray}{gray}{0.5}
\usepackage{float}
\usepackage{mathtools}

\begin{document}

\copyrightyear{2017} 
\acmYear{2017} 
\setcopyright{acmcopyright}
\acmConference{SIGIR '17}{}{August 07-11, 2017, Shinjuku, Tokyo, Japan}\acmPrice{15.00}\acmDOI{http://dx.doi.org/10.1145/3077136.3080809}
\acmISBN{978-1-4503-5022-8/17/08}

\title{End-to-End Neural Ad-hoc Ranking with Kernel Pooling}

\author{Chenyan Xiong}
\authornote{The first two authors contributed equally.}
\affiliation{
% \department{Language Technologies Institute}
\institution{Carnegie Mellon University}
%   \city{Pittsburgh} 
%   \state{PA} \country{USA}
%   \postcode{15213}
}
\email{cx@cs.cmu.edu}

\author{Zhuyun Dai$^*$}
\affiliation{
% \department{Language Technologies Institute}
\institution{Carnegie Mellon University}
%   \city{Pittsburgh} 
%   \state{PA} \country{USA}
%   \postcode{15213}
}
\email{zhuyund@cs.cmu.edu}

\author{Jamie Callan}
\affiliation{
% \department{Language Technologies Institute}
\institution{Carnegie Mellon University}
%   \city{Pittsburgh} 
%   \state{PA} \country{USA}
%   \postcode{15213}
}
\email{callan@cs.cmu.edu}

\author{Zhiyuan Liu}
\affiliation{
% \department{Department of Computer Science \\and Technology}
\institution{Tsinghua University}
%   \city{Beijing} 
%   \country{P.R. China} 
%   \postcode{100080}
}
\email{liuzy@tsinghua.edu.cn}

\author{Russell Power}
\affiliation{
\institution{Allen Institute for AI}
%   \city{Seattle} 
%   \state{WA} \country{USA}
}
\email{russellp@allenai.org}
\begin{abstract}

This paper proposes \texttt{K-NRM}, a kernel based neural model for document ranking. Given a query and a set of documents, 
\texttt{K-NRM} uses a  translation matrix that models word-level similarities via word embeddings, 
a new kernel-pooling technique that uses kernels to extract multi-level soft match features, and a learning-to-rank layer that combines those features into the final ranking score. 
The whole model is trained end-to-end.
The ranking layer learns desired feature patterns from the pairwise ranking loss. 
The kernels transfer the feature patterns into soft-match targets at each similarity level and enforce them on the translation matrix.
The word embeddings are tuned accordingly so that they can produce the desired soft matches.
Experiments on a commercial search engine's query log demonstrate the improvements of \texttt{K-NRM} over prior feature-based and neural-based states-of-the-art, and explain the source of \texttt{K-NRM}'s advantage: 
Its kernel-guided embedding encodes a similarity metric tailored for matching query words to
document words, and provides effective multi-level soft matches.
\end{abstract}

\maketitle

\section{Introduction}
\label{sec:intro}

In traditional information retrieval, queries and documents are typically represented by discrete bags-of-words,
the ranking is based on exact matches between query and document words,
and trained ranking models rely heavily on feature engineering. 
In comparison, newer neural information retrieval (neural IR) methods use continuous text embeddings, 
model the query-document relevance via soft matches, and aim to learn feature representations automatically. 
With the successes of deep learning in many related areas, neural IR has the potential to redefine the boundaries of information retrieval; however, achieving that potential has been difficult so far.

Many neural approaches use distributed representations (e.g., word2vec \cite{word2vec}), but in spite of many efforts, distributed representations have a limited history of success for document ranking.  Exact match of query words to document words is a strong signal of relevance \cite{guo2016semantic}, whereas soft-match is a weaker signal that must be used carefully.  Word2vec may consider `pittsburgh' to be similar to `boston', and `hotel' to be similar to `motel'.  However, a person searching for `pittsburgh hotel' may accept a document about `pittsburgh motel', but probably will reject a document about `boston hotel'.  How to use these soft-match signals effectively and reliably is an open problem.

This work addresses these challenges with a kernel based neural ranking model (\texttt{K-NRM}).
\texttt{K-NRM} uses distributed representations to represent query and document words.
Their similarities are used to construct a translation model.
Word pair similarities are combined by a new kernel-pooling layer that uses kernels to softly count the frequencies of word pairs at different similarity levels (soft-TF). 
The soft-TF signals are used as features in a ranking layer, which produces the final ranking score. 
All of these layers are differentiable and allow \texttt{K-NRM} to be optimized end-to-end.

% RP:
% The kernel pooling layer is key to the performance of K-NRM.  Unlike
% histogram pooling, it is
% differentiable, allowing word representations to be directly trained
% to improve document ranking (unlike word2vec).  It also preserves more
% ranking information than other common pooling mechanisms (averaging or
% max-pooling).  This not only gives the final LeTor layer more information
% for ranking, it also gives a stronger signal (via back-propagation) to the
% embedding layer, allowing for better word representations.

%
% I would cut it and give more detail about the model description above.
% If you need to keep it, maybe just keep:
%
% The entire K-NRM model is differentiable; this allows all layers to be optimized
% via gradient descent and back-propagation.

The kernels are the key to \texttt{K-NRM}'s capability. 
During learning, the kernels convert the learning-to-rank loss to requirements on soft-TF patterns, and adjust the word embeddings to produce a soft match that can better separate the relevant and irrelevant documents. 
This kernel-guided embedding learning encodes a similarity metric tailored for matching query and document. The tailored similarity metric is conveyed by the learned embeddings, which produces effective multi-level soft-matches for ad-hoc ranking.

% In the back propagation, the kernels receive the training signals from learning to rank, i.e. the desired soft-match feature patterns that better rank relevant documents over irrelevant documents, and adjust the word similarities to meet the desired soft-matches---the kernels convert the relevance labels to word embedding's training signals: word pairs that provide strong soft-match signals are moved to desired similarity levels in the embedding space, word pairs that are not useful for ranking are moved away.

%====================

% The whole model is differentiable and is trained end-to-end using back-propagation: The ranking layer learns the desired feature patterns from the ranking loss.  Kernels enforce the desired feature patterns as soft-TF budgets on the translation matrix.  The embedding model uses the budgets as guidance and adjusts the word embeddings accordingly. In this learning process, the ranking preferences are memorized and encoded by the word embeddings, soft-TF ranking features are automatically extracted by the kernels, and \texttt{K-NRM} learns how to rank documents end-to-end without requiring any feature engineering.

Extensive experiments on a commercial search engine's query log demonstrate the significant and robust advantage of \texttt{K-NRM}.
On different evaluation scenarios (in-domain, cross-domain and raw user clicks), and on different parts of the query log (head, torso, and tail), \texttt{K-NRM} outperforms both feature-based ranking and neural ranking states-of-the-art by as much as 65\%. 
\texttt{K-NRM}'s advantage is not from an unexplainable `deep learning magic', but the long-desired soft match achieved by its kernel-guided embedding learning. In our analysis, if used without the multi-level soft match or the embedding learning, the advantage of \texttt{K-NRM} quickly diminishes; while with the kernel-guided embedding learning, 
\texttt{K-NRM} successfully learns relevance-focused soft matches using its embedding and ranking layers, and the memorized ranking preferences generalize well to different testing scenarios.

The next section discusses related work. Section 3 presents the kernel-based neural ranking model. Experimental methodology is discussed in Section 4 and evaluation results are presented in Section 5. Section 6 concludes.

\section{Related Work}

Retrieval models such as query likelihood and BM25 are based on exact matching of query and document words, which limits the information available to the ranking model and may lead to problems such \textit{vocabulary mismatch}~\cite{croft2010search}.  
\emph{Statistical translation models} were an attempt to overcome this limitation. They model query-document relevance using a pre-computed \emph{translation matrix} that describes the similarities between word pairs~\cite{berger1999Information}. At query time, the ranking function considers the similarities of all query and document word pairs, allowing query words to be soft-matched to document words. The translation matrix can be calculated via mutual information in a corpus~\cite{karimzadehgan2010estimation} or using user clicks~\cite{gao2010clickthrough}.

Word pair interactions have also been modeled by \emph{word embeddings}.
Word embeddings trained from surrounding contexts, for example, word2vec~\cite{word2vec}, are considered to be the factorization of word pairs' PMI matrix~\cite{levy2015improving}. Compared to word pair similarities which are hard to learn, word embeddings provide a smooth low-level approximation of word similarities that may improve translation models~\cite{zuccon2015integrating, guo2016semantic}.

Some research has questioned whether word embeddings based on surrounding context,  such as word2vec, are suitable for ad hoc ranking.  Instead, it customizes word embeddings for search tasks. 
Nalisnick et al. propose to match query and documents using both the input and output of the embedding model, instead of only using one side of them~\cite{nalisnick2016improving}. Diaz et al. find that word embeddings trained locally on pseudo relevance feedback documents are more related to the query's information needs, and can provide better query expansion terms~\cite{diazquery}. 

Current neural ranking models fall into two groups: \emph{representation} based and \emph{interaction} based~\cite{guo2016semantic}.
The earlier focus of neural IR was mainly on \emph{representation} based models, in which the query and documents are first embedded into continuous vectors, and the ranking is calculated from their embeddings' similarity. For example,  DSSM~\cite{huang2013learning} and its convolutional version CDSSM~\cite{shen2014learning} map words to letter-tri-grams, embed query and documents using neural networks built upon the letter-tri-grams, and rank documents using their embedding similarity with the query.

The \emph{interaction} based neural models, on the other hand, learn query-document matching patterns from word-level interactions. For example, ARC-II~\cite{hu2014convolutional} and MatchPyramid~\cite{pang2016} build hierarchical Convolutional Neural Networks (CNN) on the interactions of two texts' word embeddings; they are effective in matching tweet-retweet and question-answers~\cite{hu2014convolutional}. The Deep Relevance Matching Model (DRMM) uses  pyramid pooling (histogram)~\cite{grauman2005pyramid} to summarize the word-level similarities into ranking signals~\cite{jiafeng2016deep}. 
The word level similarities are calculated from pre-trained word2vec embeddings, and the histogram counts the number of word pairs at different similarity levels. The counts are combined by a feed forward network to produce final ranking scores.
\emph{Interaction} based models and \emph{representation} based models address the ranking task from different perspectives, and can be combined for further improvements~\cite{mitra2017learning}.

This work builds upon the ideas of customizing word embeddings and the \emph{interaction} based neural models: \texttt{K-NRM} ranks documents using soft matches from query-document word interactions, and learns to encode the relevance preferences using customized word embeddings 
at the same time, which is achieved by the kernels.

\section{Kernel Based Neural Ranking}

This section presents \texttt{K-NRM}, our kernel based neural ranking model. 
We first discuss how \texttt{K-NRM} produces the ranking score for a query-document pair with their words as the sole input (ranking from scratch). Then we derive how the ranking parameters and word embeddings in \texttt{K-NRM} are trained from ranking labels (learning end-to-end).

%===== model architecture====
\begin{figure}[t]
\centering
\includegraphics[width=0.48\textwidth]{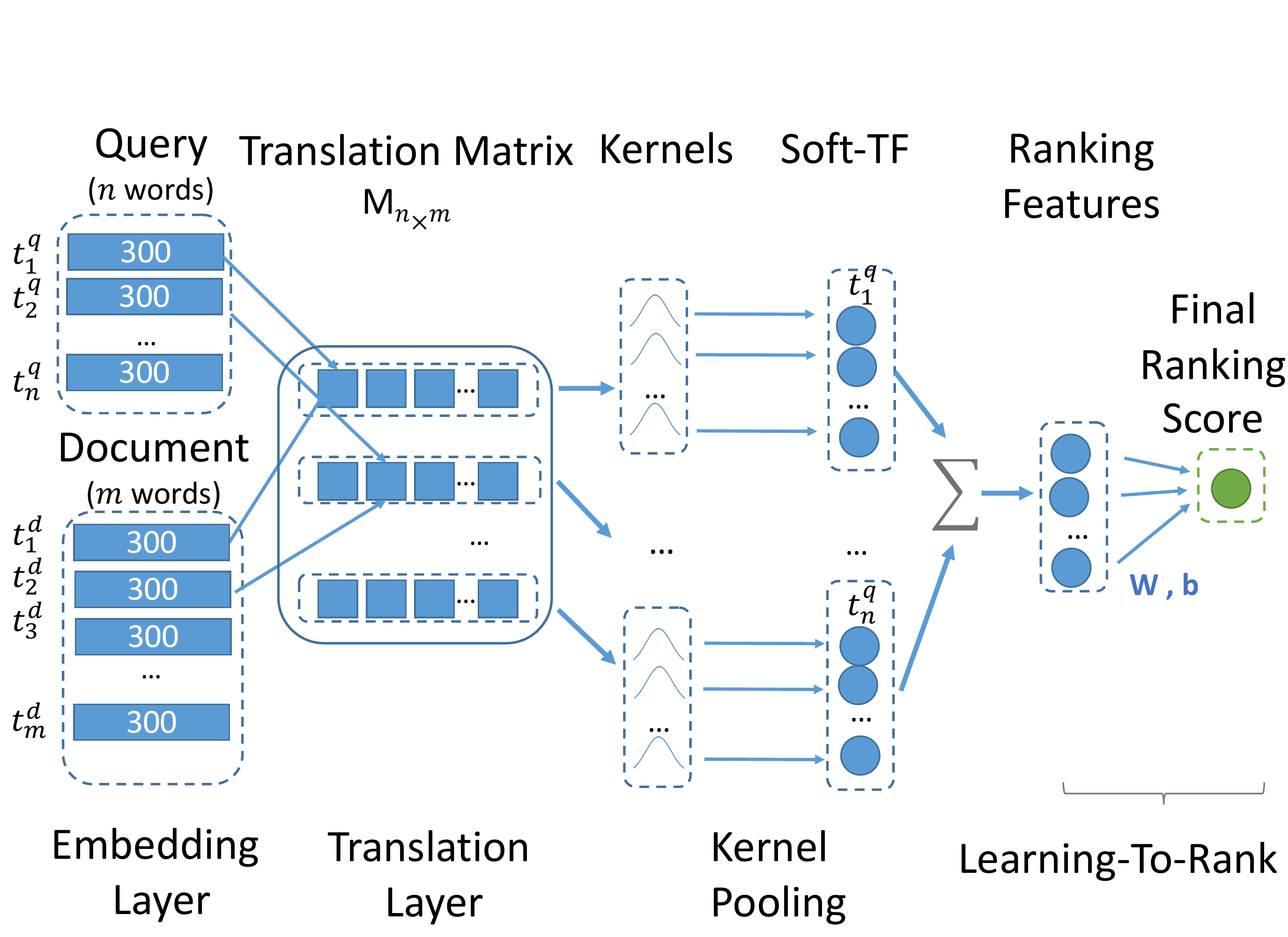}
\caption{The Architecture of \texttt{K-NRM}.
Given input query words and document words, the embedding layer maps them into distributed representations, the translation layer calculates the word-word similarities and forms the translation matrix, the kernel pooling layer generate soft-TF counts as ranking features, and the learning to rank layer combines the soft-TF to the final ranking score.
% \texttt{K-NRM} maps them into word embeddings, builds a translation matrix upon their interactions, and then uses the kernel pooling layer to count soft-TF ranking features for the leaning to rank to produce the final ranking score.
\label{fig:model}
}
\end{figure}

\subsection{Ranking from Scratch}

Given a query $q$ and a document $d$, \texttt{K-NRM} aims to generate a ranking score $f(q, d)$ only using query words  $q=\{t_1^{q},...t_i^{q}...,t_n^{q}\}$ and document words $d=\{t_1^{d},...t_j^{d}...,t_m^{d}\}$. As shown in Figure~\ref{fig:model}, \texttt{K-NRM} achieves this goal via three components: translation model, kernel-pooling, and learning to rank. 

\textbf{Translation Model:} \texttt{K-NRM} first uses an \texttt{embedding layer} to map each word $t$ to an L-dimension embedding $\vec{v}_t$:
\begin{align*}
t \Rightarrow \vec{v}_t. 
\end{align*}

Then a \texttt{translation layer} constructs a translation matrix $M$. Each element in $M$ is the embedding similarity between a query word and a document word:
\begin{align*}
M_{ij} &= \cos(\vec{v}_{t_i^{q}}, \vec{v}_{t_j^{d}}).
\end{align*}
The translation model in \texttt{K-NRM} uses word embeddings to recover the word similarities instead of trying to learn one for each word pair.
Doing so requires much fewer parameters to learn. For a vocabulary of size $|V|$ and the embedding dimension $L$, \texttt{K-NRM}'s translation model includes $|V|\times L$ embedding parameters, much fewer than learning all pairwise similarities ($|V|^2$).

\newpage
\textbf{Kernel-Pooling:} \texttt{K-NRM} then uses kernels to convert word-word interactions in the translation matrix $M$ to query-document ranking features ${\phi}(M)$:
\begin{align*}
{\phi}(M) &= \sum_{i=1}^n \log \vec{K}(M_i) \\
\vec{K}(M_i) &= \{K_1(M_i), ..., K_K(M_i)\}
\end{align*}
$\vec{K}(M_i)$ applies $K$ kernels to the $i$-th query word's row of the translation matrix, summarizing (pooling) it into a $K$-dimensional feature vector. The log-sum of each query word's feature vector forms the query-document ranking feature vector ${\phi}$. 

\begin{figure}[t]
\centering
 \begin{subfigure}[!htb]{0.235\textwidth}
\includegraphics[width=\textwidth]{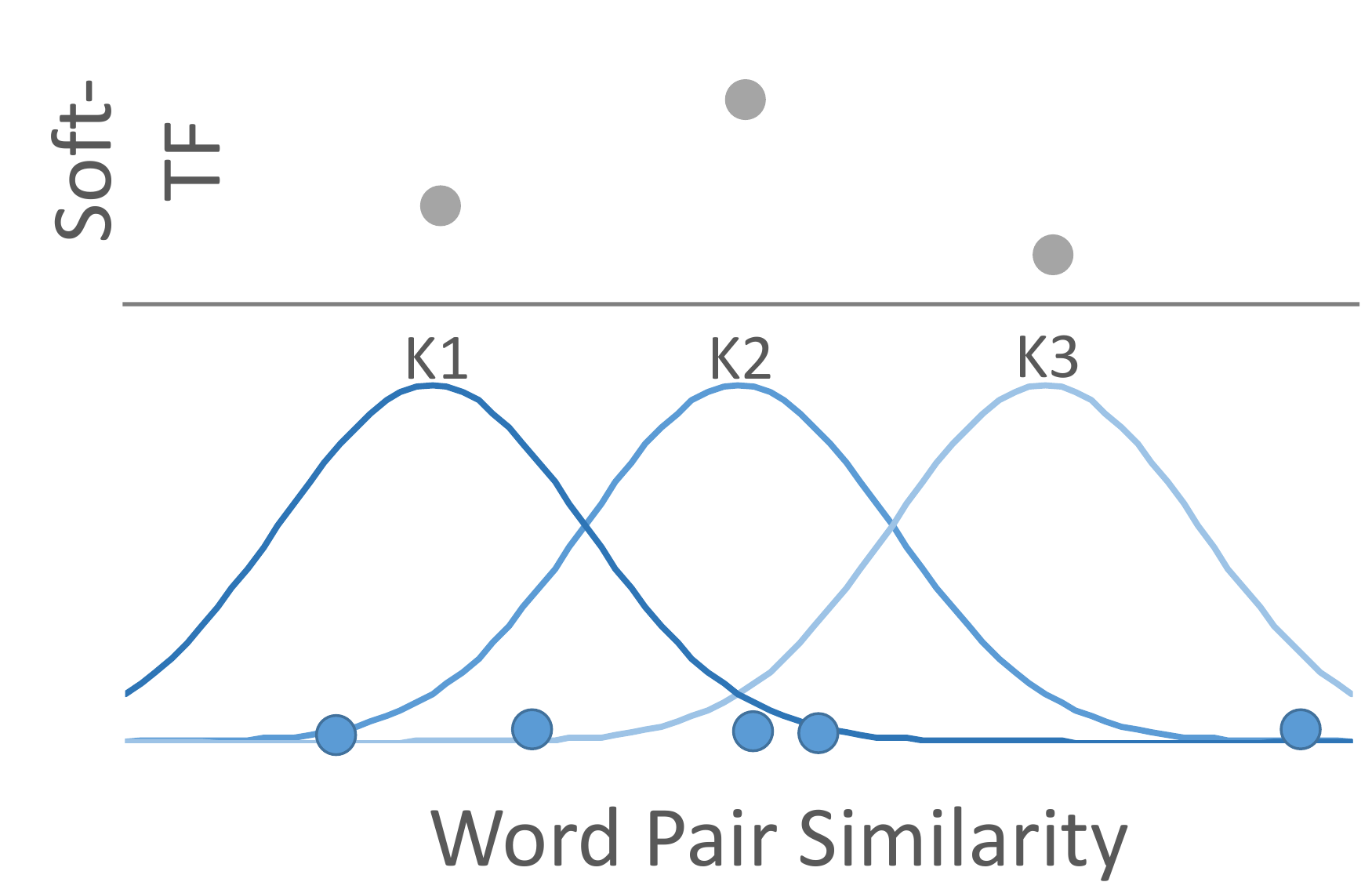}
 \caption{Ranking} \label{fig:kernel_rank}
\end{subfigure}
 \begin{subfigure}[!htb]{0.235\textwidth}
\includegraphics[width=\textwidth]{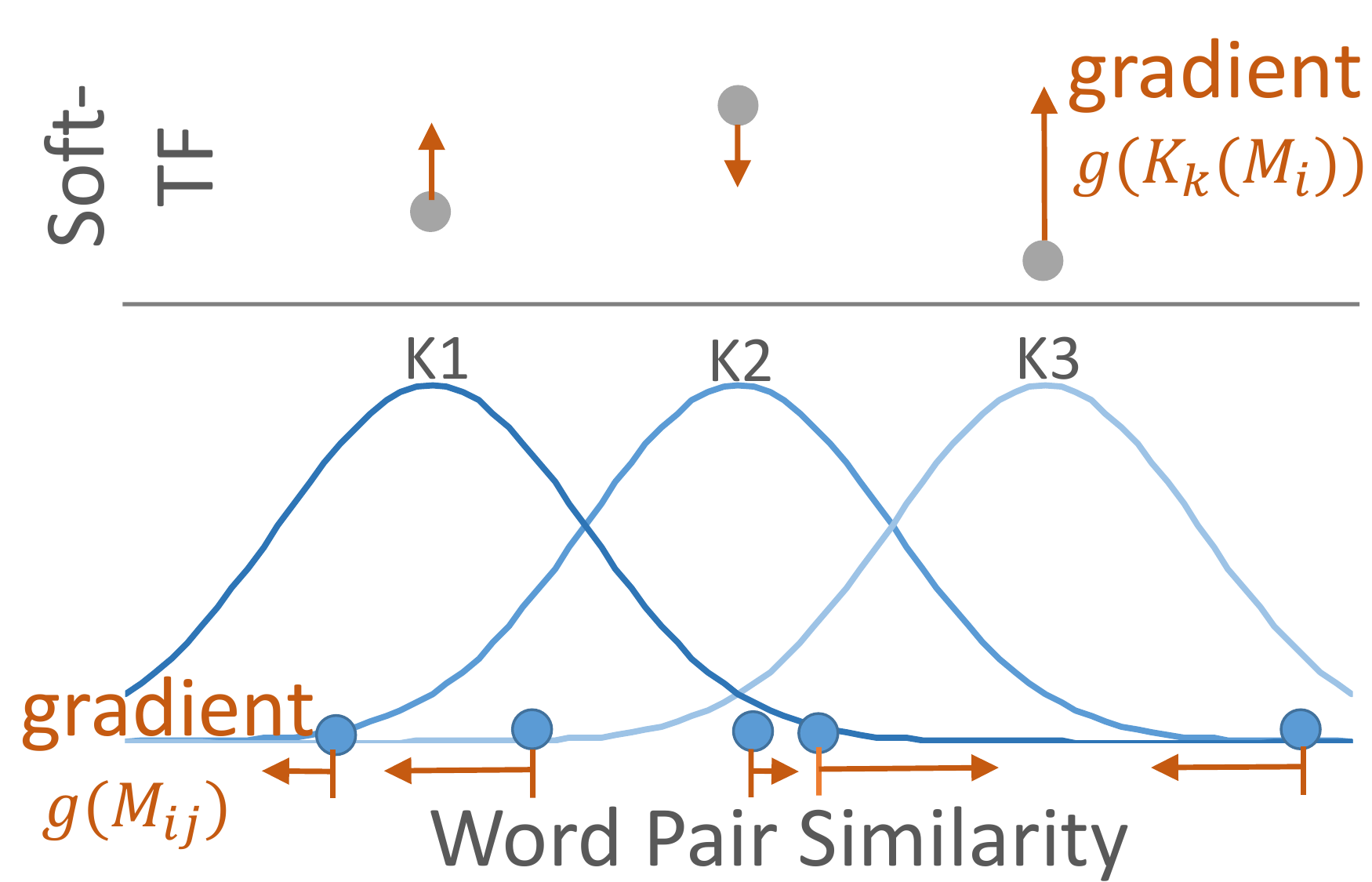}
 \caption{Learning} \label{fig:kernel_learn}
\end{subfigure}
\caption{
Illustration of kernels in the ranking (forward) process and learning (backward) process.}
\label{fig:kernel_hist}
\end{figure}

The effect of $\vec{K}$ depends on the kernel used. This work uses the RBF kernel: 
\begin{align*}
K_k(M_i) &=\sum_{j} \exp(-\frac{(M_{ij} - \mu_k)^2}{2 \sigma_k^2}).
\end{align*}
As illustrated in Figure~\ref{fig:kernel_rank}, the RBF kernel $K_k$ calculates how word pair similarities are distributed around it: the more word pairs with similarities closer to its mean $\mu_k$, the higher its value.
Kernel pooling with RBF kernels is a generalization of existing pooling techniques.
As $\sigma \rightarrow \infty$, the kernel pooling function devolves to the mean
pooling. $\mu=1$ and $\sigma \rightarrow 0$ results in a kernel that only responds to exact matches, equivalent to the TF value from sparse models.
Otherwise, the kernel functions as `soft-TF'\footnote{The RBF kernel is one of the most popular choices. 
Other kernels with similar density estimation effects can also be used, as long as they are differentiable.
For example, polynomial kernel can be used, but histograms~\cite{jiafeng2016deep} cannot as they are not differentiable.}. $\mu$ defines the similarity level that `soft-TF' focuses on; for example, a kernel with $\mu=0.5$ calculates the number of document words whose similarities to the query word are close to $0.5$. $\sigma$ defines the kernel width, or the range of its `soft-TF' count.

\textbf{Learning to Rank:} The ranking features ${\phi}(M)$ are combined by a ranking layer
to produce the final ranking score:
\begin{align*}
f(q, d) &= \text{tanh}(w^T\phi(M) + b).
\end{align*}
$w$ and $b$ are the ranking parameters to learn. $\text{tanh}()$ is the activation function.
It controls the range of ranking score to facilitate the learning process. It is rank-equivalent to a typical linear learning to rank model.

Putting every together, \texttt{K-NRM} is defined as:
\begin{align}
f(q, d) &= \text{tanh}(w^T{\phi}(M) + b) &\text{Learning to Rank} \label{eq:ltr} \\
{\phi}(M) &= \sum_{i=1}^n \log \vec{K}(M_i) &\text{Soft-TF Features} \label{eq:kp1} \\
\vec{K}(M_i) &= \{K_1(M_i), ..., K_K(M_i)\} &\text{Kernel Pooling} \label{eq:kp2}\\
K_k(M_i) &=\sum_{j} \exp(-\frac{(M_{ij} - \mu_k)^2}{2 \sigma_k^2}) &\text{RBF Kernel} \label{eq:rbf}\\
M_{ij} &= \cos(\vec{v}_{t_i^{q}}, \vec{v}_{t_j^{d}}) &\text{Translation Matrix} \label{eq:trans}\\
t &\Rightarrow \vec{v}_t. &\text{Word Embedding} \label{eq:emb}
\end{align}
Eq.~\ref{eq:trans}-\ref{eq:emb} embed query words and document words, and calculate the translation matrix. The kernels (Eq.~\ref{eq:rbf}) count the soft matches between query and document's word pairs at multiple levels, and generate $K$ soft-TF ranking features (Eq.~\ref{eq:kp1}-\ref{eq:kp2}).
Eq.~\ref{eq:ltr} is the learning to rank model. 
The ranking of \texttt{K-NRM} requires no manual features. The only input used is the query and document words. The kernels extract soft-TF ranking features from word-word interactions automatically.

\subsection{Learning End-to-End}
\label{sec:model_optimization}
The \textbf{training} of \texttt{K-NRM} uses the pairwise learning to rank loss:
\begin{align}
l(w, b, \mathcal{V}) = \sum_q \sum_{d^+, d^- \in D_q^{+, -}} max(0, 1 - f(q, d^+) + f(q, d^-)). \label{eq:pair_loss}
\end{align}
$D_q^{+, -}$ are the pairwise preferences from the ground truth: $d^+$ ranks higher than $d^-$.
The parameters to learn include the ranking parameters $w, b$, and the word embeddings $\mathcal{V}$.

The parameters are optimized using back propagation (BP) through the neural network. 
Starting from the ranking loss, the gradients are first propagated to the learning-to-rank part (Eq.~\ref{eq:ltr}) and update the ranking parameters ($w, b$), the kernels pass the gradients to the word similarities (Eq.~\ref{eq:kp1}-\ref{eq:rbf}), and then to the embeddings (Eq.~\ref{eq:trans}).

\textbf{Back propagations through the kernels:} 
The embeddings contain millions of parameters $\mathcal{V}$ and are the main capacity of the model. The learning of the embeddings is guided by the kernels.  

The back propagation first applies gradients from the loss function (Eq.~\ref{eq:pair_loss}) to the ranking score $f(q, d)$, to increase it (for $d^+$) or decrease it (for $d^-$); 
the gradients are propagated through Eq.~\ref{eq:ltr} to the feature vector $\phi(M)$, and then through Eq.~\ref{eq:kp1} to the the kernel scores $\vec{K}(M_i)$. 
The resulted $g(\vec{K}(M_i))$ is a $K$ dimensional vector:
\begin{align*}
g(\vec{K}(M_i)) &= \{g(K_1(M_i)),...,g(K_K(M_i)\}.
\end{align*}
Its each dimension $g(K_k(M_i))$ is jointly defined by the ranking score's gradients and the ranking parameters. It adjusts the corresponding kernel's score up or down to better separate the relevant document ($d^+$) from the irrelevant one ($d^-$).

The kernels spread the gradient to word similarities in the translation matrix $M_{ij}$, through Eq.~\ref{eq:rbf}: 
\begin{align}
g(M_{ij}) &=  \sum_{k=1}^K \frac{g(K_k(M_i)) \times \sigma_k^2}{(\mu_k - M_{ij}) \exp(\frac{(M_{ij} - \mu_k)^2}{-2\sigma_k^2}) }. \label{eq:grbf}
\end{align}

% \noindent
% Intuitively, $g(K_k(M_i))$ is the updates applied to the $k$-th kernel in the back propagation, i.e. to form better ranking feature patterns that better separate relevant documents from irrelevant ones. 
% The kernels propagate $g(K_k(M_i))$ to word pair similarities $M_{i, j}$.
The kernel-guided embedding learning process is illustrated in Figure~\ref{fig:kernel_learn}.
% As shown in Figure~\ref{fig:kernel_learn}, 
A kernel pulls the word similarities closer to its $\mu$ to increase its soft-TF count, or pushes the word pairs away to reduce it, based on the gradients received in the back-propagation. 
The \emph{strength} of the force also depends on the the kernel's width $\sigma_k$ and the word pair's distance to $\mu_k$: approximately, the wider the kernel is (bigger $\sigma_k$), and the closer the word pair's similarity to $\mu_k$, the stronger the force is (Eq. \ref{eq:grbf}). 
The gradient a word pair's similarity received, $g(M_{ij})$,  is the combination of the forces from all $K$ kernels.
% Each word pair receives $K$ different forces from the $K$ kernels. The combination of the forces, $g(M_{ij})$, decides the movement of the word pair.

%$g(M(i, j))$ is the adjustment of the word pair similarities to fulfill the request posted by  $g(K_k(M_i))$. 
%The \emph{direction} of $g(M(i ,j))$ is defined by the direction of $g(K_k(M_i)$ and the location of $M(i, j)$ respect to $\mu_k$ ($\mu_k - M(i ,j)$). As shown in Figure~\ref{fig:kernel_learn}, the kernel pulls the word similarities closer to its $\mu$ to increase its soft-TF, or pushes the word pairs away to reduce it. The \emph{strength} of the pull/push is defined by the kernel's width $\sigma_k$ and the distance to $\mu_k$: intuitively, the wider the kernel is (bigger $\sigma_k$), the closer the word pair's similarity to $\mu_k$, the stronger the force is.

The word embedding model receives $g(M_{ij})$ and updates the embeddings accordingly. 
Intuitively, the learned word embeddings are aligned to form multi-level soft-TF patterns that can separate the relevant documents from the irrelevant ones in training, and the learned embedding parameters $\mathcal{V}$ memorize this information.
% memorize this information using embedding parameter $\mathcal{V}$.
When testing, \texttt{K-NRM} extracts soft-TF features from the learned word embeddings using the kernels and produces the final ranking score using the ranking layer.

\section{Experimental Methodology}
This section describes our experimental methods and materials.

\subsection{Dataset}
Our experiments use a query log sampled from search logs of Sogou.com, a major Chinese commercial search engine. The sample contains 35 million search sessions with $96,229$ distinct queries. The query log includes queries, displayed documents, user clicks, and dwell times. Each query has an average of $12$ documents displayed.  As the results come from a commercial search engine, the returned documents tend to be of very high quality.

The primary testing queries were $1,000$ queries sampled from head queries that appeared more than $1,000$ times in the query log. Most of our evaluation focuses on the head queries; we use tail query performance to evaluate model robustness.
The remaining queries were used to train the neural models.  Table~\ref{tab:dataset} provides summary statistics for the training and testing portions of the search log.

The query log contains only document titles and URLs. The full texts of \emph{testing} documents were crawled and parsed using Boilerpipe~\cite{boilerpipe} for our word-based baselines (described in Section \ref{sec:baselines}). 
Chinese text was segmented using the open source software ICTCLAS~\cite{zhang2003hhmm}. After segmentation, documents are treated as sequences of words (as with English documents).

\begin{table}[t]
\centering
\caption{Training and testing dataset characteristics.}
\label{tab:dataset}
\begin{tabular}{l|r|r}
\hline
& \textbf{Training} & \textbf{Testing} \\ \hline
{Queries}                                                                                          & 95,229    & 1,000
\\ 
{Documents Per Query}                                                                                        & 12.17 & 30.50 \\
{Search Sessions} & 31,201,876 & 4,103,230 \\ 
{Vocabulary Size}                                                                                  & 165,877   & 19,079\\ \hline
\end{tabular}
\end{table}

\begin{table}[t]
\centering
\caption{Testing Scenarios. DCTR Scores are inferred by DCTR click model~\cite{chuklin2015click}. TACM Scores are inferred by TACM click model~\cite{liu2016time}. Raw Clicks use the sole click in a session as the positive label. The label distribution is the number of relevance labels from 0-4 from left to right, if applicable.
\label{tab:test}
}
\begin{tabular}{l|l|l}
\hline
\textbf{Condition} & \textbf{Label} & \textbf{Label Distribution}\\ \hline
\textbf{Testing-SAME} & DCTR Scores & 70\%, 19.6\%, 9.8\%, 1.3\%, 1.1\%  \\
\textbf{Testing-DIFF} &  TACM Scores &  79\%, 14.7\%, 4.6\%, 0.9\%, 0.9\% \\
\textbf{Testing-RAW} &  Raw Clicks & 2,349,280 clicks\\ \hline
\end{tabular}
\end{table}

\subsection{Relevance Labels and Evaluation Scenarios}
\label{sec:scenarios}
Neural models like \texttt{K-NRM} and \texttt{CDSSM} require a large amount of training
data.  Acquiring a sufficient number of manual training labels outside
of a large organization would be cost-prohibitive.  User click data, on the other hand, is easy
to acquire and prior research has shown that it can accurately predict manual labels.  For our experiments
\textbf{training labels} were generated based on user clicks from the training sessions. 

There is a large amount of prior research on building \textit{click models} to model user behavior and to infer reliable relevance signals from clicks~\cite{chuklin2015click}. This work uses one of the simplest click models, DCTR, 
to generate relevance scores from user clicks~\cite{chuklin2015click}. DCTR calculates the relevance scores 
of a query-document pair based on their click through rates. Despite being extremely simple, it performs rather 
well and is a widely used baseline~\cite{chuklin2015click}. 
Relevance scores from DCTR are then used to generate preference pairs to train our models.

The \textbf{testing labels} were also estimated from the click log, as manual relevance judgments were not made available to us.
\textit{Note that there was no overlap between training queries and testing queries.} 

\textbf{Testing-SAME} infers relevance labels using DCTR, the same click model used for training. This setting evaluates the ranking model's ability to fit user preferences (click through rates).

\textbf{Testing-DIFF} infers relevance scores using TACM~\cite{liu2016time}, a state-of-the-art click model. TACM is a more sophisticated model and uses both clicks and dwell times.
On an earlier sample of Sogou's query log, the TACM labels aligned extremely well with expert annotations: when evaluated against manual labels, TACM achieved an NDCG@5 of up to 0.97~\cite{liu2016time}.
This is substantially higher than the agreement between the manual labels generated by the authors for a sample of queries. 
This precision makes TACM's inferred scores a good approximation of expert labels, and Testing-DIFF is expected to produce evaluation results similar to expert labels.

\textbf{Testing-RAW} is the simplest click model.
Following the cascade assumption~\cite{chuklin2015click}, we treat the clicked document in each single-click session as a relevant document, and test whether the model can put it at the top of the ranking. Testing-Raw only uses single-click sessions ( $57\%$ of the testing sessions are single-click sessions).  Testing-RAW is a conservative setting that uses \emph{raw} user feedback. It eliminates the influence of click models in testing, and evaluates the ranking model's ability to overcome possible disturbances from the click models.  

The three testing scenarios are listed in Table~\ref{tab:test}.
Following TREC methodology, the Testing-SAME and Testing-DIFF's inferred relevance scores were mapped to 5 relevance grades.  Thresholds were chosen so that our relevance grades have the same distribution as TREC Web Track 2009-2012 qrels.

Search quality was measured using NDCG at depths $\{1, 3, 10\}$ for Testing-SAME and Testing-DIFF. We focused on early ranking positions that are more important for commercial search engines.
Testing-RAW was evaluated by mean reciprocal rank (MRR) as there is only one relevant document per query.
Statistical significance was tested using the permutation test with p $<0.05$.

\subsection{Baselines}
\label{sec:baselines}

Our baselines include both traditional word-based ranking models as well as more recent neural ranking models. 

\textbf{Word-based baselines} include \texttt{BM25} and language models with Dirichlet smoothing (\texttt{Lm}). These unsupervised retrieval methods were applied on the full text of candidate documents, and used to re-rank them. We found that these methods performed better on full text than on titles. Full text default parameters were used.

Feature-based learning to rank baselines include \texttt{RankSVM}\footnote{https://www.cs.cornell.edu/people/tj/svm\_light/svm\_rank.html}, a state-of-the-art pairwise ranker, and coordinate ascent~\cite{metzler2007linear}
(\texttt{Coor}-\texttt{Ascent}\footnote{https://sourceforge.net/p/lemur/wiki/RankLib/}), a state-of-the-art listwise ranker.
They use typical word-based features: Boolean AND; Boolean OR; Coordinate match; TF-IDF; BM25; language models with no smoothing, Dirichlet smoothing, JM smoothing and two-way smoothing; and bias. All features were applied to the document title and body. The parameters of the retrieval models used in feature extraction are kept default.

\begin{table}[t]
\centering
\caption{The number of parameters and the word embeddings used by baselines and \texttt{K-NRM}. `--' indicates not applicable, e.g. unsupervised methods have no parameters, and word-based methods do not use embeddings.}
\label{tab:param}
\begin{tabular}{l|c|c}
\hline
\textbf{Method}           & \textbf{Number of Parameters}   & \textbf{Embedding}\\ \hline
\texttt{Lm}, \texttt{BM25}            & --               & --    \\
\texttt{RankSVM}            & 21               & --    \\
\texttt{Coor-Ascent}            & 21               & --    \\
\texttt{Trans}            & --     & word2vec    \\
\texttt{DRMM}             & 161              & word2vec    \\
\texttt{CDSSM}            & 10,877,657  &     --    \\ 
\texttt{K-NRM}        & 49,763,110       & end-to-end    \\ 
\hline

\end{tabular}
\end{table}

\textbf{Neural ranking baselines} include \texttt{DRMM}~\cite{jiafeng2016deep}, \texttt{CDSSM}~\cite{shen2014latent},  and a simple embedding-based translation model, \texttt{Trans}.

\texttt{DRMM} is the state-of-the-art interaction based neural ranking model~\cite{jiafeng2016deep}. 
It performs histogram pooling on the embedding based translation matrix and uses the binned soft-TF as the input to a ranking neural network. The embeddings used are pre-trained via word2vec~\cite{word2vec} because the histograms are not differentiable and prohibit end-to-end learning.
We implemented the best variant, $\texttt{DRMM}_{\texttt {LCH}\times \texttt{IDF}}$. The pre-trained embeddings were obtained by applying the skip-gram method from word2vec on our training corpus (document titles displayed in training sessions).

\texttt{CDSSM}~\cite{shen2014learning} is the convolutional version of DSSM~\cite{huang2013learning}. 
\texttt{CDSSM} maps English words to letter-tri-grams using a word-hashing technique, and uses Convolutional Neural Networks to build representations of the query and document upon the letter-tri-grams. It is a state-of-the-art representation based neural ranking model.
We implemented \texttt{CDSSM} in Chinese by convolving over Chinese characters.  (Chinese characters can be considered as similar to English letter-tri-grams with respect to word meaning).
\texttt{CDSSM} is also an end-to-end model, but uses discrete letter-tri-grams/Chinese characters instead of word embeddings.

\texttt{Trans} is an unsupervised embedding based translation model.
Its translation matrix is calculated by the cosine similarity of word embeddings from the same word2vec used in \texttt{DRMM}, and then averaged to the query-document ranking score.

\textbf{Baseline Settings:}
\texttt{RankSVM} uses a linear kernel and the hyper-parameter C was selected in the development fold of the cross validation from the range $[0.0001, 10]$. 

Recommended settings from RankLib were used for \texttt{Coor-Ascent}.

We obtained the body texts of testing documents from the new Sogou-T corpus~\cite{Sogou-T} or crawled them directly. The body texts were used by all word-based baselines. Neural ranking baselines and \texttt{K-NRM} used only titles for training and testing, as the coverage of Sogou-T on the training documents is low and the training documents could not be crawled given resource constraints.  

For all baselines, the most optimistic choices were made: feature-based methods (\texttt{RankSVM} and \texttt{Coor-Ascent}) were 
trained using 10-fold cross-validation on the \emph{testing} set and use both document title and body texts.
The neural models were trained on the training set with the same settings as \texttt{K-NRM}, and only use document titles (they still perform better than only using the testing data).
Theoretically, this gives the sparse models a slight performance advantage as their training and testing data were drawn from the same distribution.

\begin{table*}[h]
\centering
\caption{
Ranking accuracy of \texttt{K-NRM} and baseline methods.
Relative performances compared with \texttt{Coor-Ascent} are in percentages. 
\textbf{W}in/\textbf{T}ie/\textbf{L}oss are the number of queries improved, unchanged, or hurt, compared to \texttt{Coor-Ascent} on NDCG@10. $\dagger, \ddagger, \mathsection$, $\mathparagraph$  indicate statistically significant improvements over \texttt{Coor-Ascent}$^\dagger$, \texttt{Trans}$^\ddagger$, \texttt{DRMM}$^\mathsection$ and \texttt{CDSSM}$^\mathparagraph$, respectively.
}

\begin{subtable}{1\textwidth}
\centering
\caption{Testing-SAME. Testing labels are inferred by the same click model (DCTR) as the training labels used by neural models.
\label{tab:ranking_res_dctr}
}
\begin{tabular}{l|lr|lr|lr|c}
\hline
% ------- DCTR ---------------
\bf{Method} &
\multicolumn{2}{c|}{\bf{NDCG@1}} &
\multicolumn{2}{c|}{\bf{NDCG@3}} &
\multicolumn{2}{c|}{\bf{NDCG@10}} & 
\bf{W/T/L}
\\ \hline
% \texttt{TF-IDF}
%  & ${0.1332}$ & $ -2.49\%  $ 

%  & ${0.1668}$ & $ -12.29\%  $ 

%  & ${0.2860}$ & $ -9.19\%  $ 

% & 327/151/429

% \\
\texttt{Lm}
 & ${0.1261}$ & $ -20.89\%  $ 

 & ${0.1648}$ & $ -26.46\%  $ 

 & ${0.2821}$ & $ -20.45\%  $ 

& 293/116/498

\\
\texttt{BM25}
 & ${0.1422}$ & $ -10.79\%  $ 

 & ${0.1757}$ & $ -21.60\%  $ 

 & ${0.2868}$ & $ -10.14\%  $ 

& 299/125/483

\\ \hline
\texttt{RankSVM}
 & ${0.1457}$ & $ -8.59\%  $ 

 & ${0.1905}$ & $ -14.99\%   $ 

 & ${0.3087}$ & $ -12.97\%  $ 

& 371/151/385
\\
\texttt{Coor-Ascent}
 & ${0.1594}^{{\ddagger \mathsection \mathparagraph } }$ &  -- 

 & ${0.2241}^{{\ddagger \mathsection \mathparagraph }}$ &  --  

 & ${0.3547}^{{\ddagger \mathsection \mathparagraph }}$ &  --  

& --/--/--

\\ \hline
\texttt{Trans}
 & ${0.1347}$ & $ -15.50\%  $ 

 & ${0.1852}$ & $ -17.36\%  $ 

 & ${0.3147}$ & $ -11.28\%  $ 

& 318/140/449

\\
\texttt{DRMM}
 & $0.1366$ & $-14.30\%$ 
 & $0.1902$ & $-15.13\%$ 
 & $0.3150$ & $-11.20\%$ 
& 318/132/457
\\ 
\texttt{CDSSM}
 & ${0.1441}$ & $ -9.59\%  $ 

 & ${0.2014}$ & $ -10.13\%  $ 

 & ${0.3329}^{\ddagger \mathsection}$ & $ -6.14\%  $ 

& 341/149/417

\\ \hline
\texttt{K-NRM}
 & $\bf{0.2642}^{\dagger \ddagger \mathsection \mathparagraph }$ & $ +65.75\%  $ 

 & $\bf{0.3210}^{\dagger \ddagger \mathsection \mathparagraph }$ & $ +43.25\%  $ 

 & $\bf{0.4277}^{\dagger \ddagger \mathsection \mathparagraph }$ & $ +20.58\%  $ 

& 447/153/307

\\ \hline
\\
\end{tabular}

\end{subtable}

% ------- TACM ---------------
\begin{subtable}{1\textwidth}
\centering
\caption{
Testing-DIFF. Testing labels are inferred by a different click model, TACM, which approximates expert labels very well~\cite{liu2016time}.
\label{tab:ranking_res_tacm}
}
\begin{tabular}{l|lr|lr|lr|c}
 \hline
\bf{Method} &
\multicolumn{2}{c|}{\bf{NDCG@1}} &
\multicolumn{2}{c|}{\bf{NDCG@3}} &
\multicolumn{2}{c|}{\bf{NDCG@10}} & 
\bf{W/T/L}
\\ \hline
\texttt{Lm}
 & ${0.1852}$ & $ -11.34\%  $ 

 & ${0.1989}$ & $ -17.23\%  $ 

 & ${0.3270}$ & $ -13.38\%  $ 

& 369/50/513

\\
\texttt{BM25}
 & ${0.1631}$ & $ -21.92\%  $ 

 & ${0.1894}$ & $ -21.18\%  $ 

 & ${0.3254}$ & $ -13.81\%  $ 

& 349/53/530

\\\hline
\texttt{RankSVM}
 & ${0.1700}$ & $ -18.62\%   $ 

 & ${0.2036}$ & $ -15.27\%   $ 

 & ${0.3519}$ & $ -6.78\%   $ 

& 380/75/477

\\
\texttt{Coor-Ascent}
 & ${0.2089}^{\ddagger \mathparagraph}$ & --  
 & $0.2403^{{\ddagger }}$ & --  
 & $0.3775^{{\ddagger \mathparagraph }}$ & --  
 & --/--/--
\\ \hline
\texttt{Trans}
 & ${0.1874}$ & $ -10.29\%  $ 

 & ${0.2127}$ & $ -11.50\%  $ 

 & ${0.3454}$ & $ -8.51\%  $ 

& 385/68/479

\\ 

\texttt{DRMM}
 & ${0.2068}$ & $ -1.00\%  $ 

 & ${0.2491}^{\ddagger }$ & $ +3.67\%  $ 

 & ${0.3809}^{\ddagger \mathparagraph }$ & $ +0.91\%  $ 

& 430/66/436

\\

\texttt{CDSSM}
 & ${0.1846}$ & $ -10.77\%  $ 

 & ${0.2358}^{\ddagger }$ & $ -1.86\%  $ 

 & ${0.3557}$ & $ -5.79\%  $ 

& 391/65/476

\\ \hline
\texttt{K-NRM}
 & $\bf{0.2984}^{\dagger \ddagger \mathsection \mathparagraph }$ & $ +42.84\%  $ 

 & $\bf{0.3092}^{\dagger \ddagger \mathsection \mathparagraph }$ & $ +28.26\%  $ 

 & $\bf{0.4201}^{\dagger \ddagger \mathsection \mathparagraph }$ & $ +11.28\%  $ 

& 474/63/395

\\ \hline

\end{tabular}
\end{subtable}

\end{table*}

\subsection{Implementation Details}
This section describes the configurations of our \texttt{K-NRM} model. 

\textbf{Model training} was done on the full training data as in Table~\ref{tab:dataset}, with training labels inferred by DCTR, as described in Section~\ref{sec:scenarios}. 
%only document titles are used in training and testing, for \texttt{K-NRM} and all neural ranking baselines.

The \textbf{embedding layer} used $300$ dimensions. The vocabulary size of our training data was $165,877$. The embedding layer was initialized with the word2vec trained on our training corpus.

The \textbf{kernel pooling layer} had $K=11$ kernels.
One kernel harvests exact matches, using $\mu_0=1.0$ and $\sigma=10^{-3}$. $\mu$ of the other 10 kernels is spaced evenly in the range of $[-1,1]$, that is $\mu_1=0.9, \mu_2=0.7,...,\mu_{10} = -0.9$. 
These kernels can be viewed as 10 soft-TF bins. $\sigma$ is set to $0.1$.
The effects of varying $\sigma$ are studied in Section \ref{sec:sigma}.

\textbf{Model optimization} used the Adam optimizer, with batch size $16$, learning rate $=0.001$ and $\epsilon=1e-5$. Early stopping was used with a patience of 5 epochs.
We implemented our model using TensorFlow.  The model training took about 50 milliseconds per batch, and converged in 12 hours on an AWS GPU machine.

Table~\ref{tab:param} summarizes the number of parameters used by the baselines and \texttt{K-NRM}. Word2vec refers to pre-trained word embeddings using skip-gram on the training corpus. End-to-end means that the embeddings were trained together with the ranking model. 

\texttt{CDSSM} learns hundreds of convolution filters on Chinese \emph{characters}, thus has millions of parameters.
\texttt{K-NRM}'s parameter space is even larger as it learns an embedding for every Chinese \emph{word}. Models with more parameters in general are expected to fit better but may also require more training data to avoid overfitting.

\section{Evaluation Results}

Our experiments investigated  \texttt{K-NRM}'s effectiveness, as well as
its behavior on tail queries, with less training data, and with different kernel widths.

\begin{table}[t]
\centering
\caption{
Ranking performance on Testing-RAW. MRR evaluates the mean reciprocal rank of clicked documents in single-click sessions. 
Relative performance in the percentages and W(in)/T(ie)/L(oss) are compared to \texttt{Coor-Ascent}. $\dagger, \ddagger, \mathsection$, $\mathparagraph$  indicate statistically significant improvements over \texttt{Coor-Ascent}$^\dagger$, \texttt{Trans}$^\ddagger$, \texttt{DRMM}$^\mathsection$ and \texttt{CDSSM}$^\mathparagraph$, respectively.
\label{tab:ranking_res_raw}
}
\begin{tabular}{l|lr|c}
 \hline
\bf{Method} &
\multicolumn{2}{c|}{\bf{MRR}} & 
\bf{W/T/L} \\ \hline
\texttt{Lm}
  & ${0.2193}$ & $ -9.19\%  $ 

& 416/09/511

\\
\texttt{BM25}
 & ${0.2280}$ & $ -5.57\%  $ 

& 456/07/473

\\ \hline
\texttt{RankSVM}
 & ${0.2241}$ & $ -7.20\%  $ 

& 450/78/473

\\
\texttt{Coor-Ascent}
 & $0.2415^{\ddagger }$ & --  & --/--/--/\\
 \hline
\texttt{Trans}
 & ${0.2181}$ & $ -9.67\%  $ 

& 406/08/522
\\
\texttt{DRMM}
 & ${0.2335}^{\ddagger }$ & $ -3.29\%  $ 

& 419/12/505

\\
\texttt{CDSSM}
 & ${0.2321}^{\ddagger }$ & $ -3.90\%  $ 

& 405/11/520
\\ \hline
\texttt{K-NRM}
 & $\bf{0.3379}^{\dagger \ddagger \mathsection \mathparagraph }$ & $ +39.92\%  $ 

& 507/05/424
\\ \hline

\end{tabular}
\end{table}

\begin{table*}[t!]
\centering
\caption{The ranking performances of several \texttt{K-NRM} variants.
Relative performances and statistical significances are all compared with \texttt{K-NRM}'s \texttt{full model}. $\dagger, \ddagger,\mathsection, \mathparagraph, $ and $*$ indicate statistically significant improvements over \texttt{K-NRM}'s variants of \texttt{exact-match}$^\dagger$, \texttt{word2vec}$^\ddagger$, \texttt{click2vec}$^\mathsection$, \texttt{max-pool}$^\mathparagraph$, and \texttt{mean-pool}$^*$, respectively. 
\label{tab:variant}
}
\begin{tabular}{l|lr|lr||lr|lr||lr}
 \hline 
 & 
 \multicolumn{4}{c||}{\bf{Testing-SAME}} &
\multicolumn{4}{c||}{\bf{Testing-DIFF}} &
\multicolumn{2}{c}{\textbf{Testing-RAW}}

\\ \hline
\bf{K-NRM Variant} &
\multicolumn{2}{c|}{\bf{NDCG@1}} &
\multicolumn{2}{c||}{\bf{NDCG@10}} &
\multicolumn{2}{c|}{\bf{NDCG@1}} &
\multicolumn{2}{c||}{\bf{NDCG@10}} &
\multicolumn{2}{c}{\bf{MRR}}
\\ \hline

 \texttt{exact-match}
 & $0.1351$ & $-49\%$

 & $0.2943$ & $-31\%$
 & $0.1788$ & $-40\%$

 & ${0.3460}^{\mathparagraph }$ & $-18\%$
 & $0.2147$ & $-37\%$
\\ \hline

 \texttt{word2vec}
 & ${0.1529}^{\dagger}$ & $-42\%$

 & ${0.3223}^{\dagger \mathparagraph}$ & $-24\%$
 & ${0.2160}^{\dagger \mathparagraph }$ & $-27\%$

 & ${0.3811}^{\dagger \mathparagraph}$ &  $-10\%$
 & ${0.2427}^{\dagger \mathparagraph}$ &  $-28\%$
\\ 

\texttt{click2vec}
 & ${0.1600}$ & $ -39\%  $ 

 & ${0.3790}^{\dagger \ddagger \mathparagraph}$ & $ -11\%  $ 
 & ${0.2314}^{\dagger \mathparagraph}$ & $ -23\%  $ 

 & ${0.4002}^{\dagger \ddagger \mathparagraph *}$ & $ -4\%  $ 
 & ${0.2667}^{\dagger \ddagger \mathparagraph}$ & $-21\%$

\\  \hline

  \texttt{max-pool} 
 & $0.1413$ &  $-47\%$

 & $0.2979$ & $-30\% $
 
 & $0.1607$ &  $-46\%$

 & $0.3334$ & $-21\%$
 & $0.2260$ & $-33\%$
 \\ 
 
 \texttt{mean-pool}
 & $0.2297^{\dagger \ddagger \mathsection \mathparagraph}$ & $-13\%$

 & $0.3614^{\dagger \ddagger \mathsection \mathparagraph}$ & $-16\% $
  & ${0.2424}^{\dagger \mathparagraph}$ & $-19\%$

 & $0.3787^{\dagger \mathparagraph}$ & $-10\%$
 & ${0.2714}^{\dagger \ddagger \mathparagraph}$ & $-20\%$
 \\
 
 \hline

\texttt{full model} 
 & $\bf{0.2642}^{\dagger \ddagger \mathsection \mathparagraph *}$ & --

 & $\bf{0.4277}^{\dagger \ddagger \mathsection \mathparagraph *}$ & --
 
 & $\bf{0.2984}^{\dagger \ddagger \mathsection \mathparagraph *}$ & --

 & $\bf{0.4201}^{\dagger \ddagger \mathparagraph *}$ & --
 & $\bf{0.3379}^{\dagger \ddagger \mathsection \mathparagraph *}$ & --
\\ \hline

\end{tabular}
\end{table*}

\subsection{Ranking Accuracy}

% ============ overall performance ================
Tables \ref{tab:ranking_res_dctr},  \ref{tab:ranking_res_tacm} and \ref{tab:ranking_res_raw} show the ranking accuracy of 
\texttt{K-NRM} and our baselines under three conditions.

\textbf{Testing-SAME} (Table \ref{tab:ranking_res_dctr}) evaluates the model's ability to fit user preferences when trained and evaluated on labels generated by the same click model (DCTR).
\texttt{K-NRM} outperforms word-based baselines by over $65\%$ on NDCG@1, and over $20\%$ on NDCG@10. The improvements over neural ranking models are even bigger: On NDCG@1 the margin between \texttt{K-NRM} and the next best neural model is $83\%$, and on NDCG@10 it is $28\%$.

\textbf{Testing-DIFF} (Table \ref{tab:ranking_res_tacm}) evaluates the model's relevance matching performance by testing on TACM inferred relevance labels, a good approximation of expert labels.
Because the training and testing labels were generated by different click models, Testing-DIFF challenges each model's ability to fit the underlying relevance signals despite perturbations caused by differing click model biases. Neural models with larger parameter spaces tend to be more vulnerable to this domain difference: \texttt{CDSSM} actually performs worse than \texttt{DRMM}, despite using thousands times more parameters. However, \texttt{K-NRM} demonstrates its robustness and is able to outperform all baselines by more than $40\%$ on NDCG@1 and $10\%$ on NDCG@10.

\textbf{Testing-RAW} (Table \ref{tab:ranking_res_raw}) evaluates each model's effectiveness directly by user clicks. It tests how well the model ranks the most satisfying document (the only one clicked) in each session. \texttt{K-NRM} improves MRR from $0.2415$ (\texttt{Coor-Ascent}) to $0.3379$. This difference is equal to moving the clicked document's from rank $4$ to rank $3$. The MRR and NDCG@1 improvements demonstrate \texttt{K-NRM}'s precision oriented property---its biggest advantage is on the earliest ranking positions. This characteristic aligns with \texttt{K-NRM}'s potential role in web search engines: as a sophisticate re-ranker, \texttt{K-NRM} is most possibly used at the final ranking stage, in which the first relevant document's ranking position is the most important.

The two neural ranking baselines \texttt{DRMM} and \texttt{CDSSM} perform similarly in all three testing scenarios.  The \emph{interaction} based model, \texttt{DRMM}, is more robust to click model biases and performs slightly better on Testing-DIFF, while the \emph{representation} based model, \texttt{CDSSM}, performs slightly better on Testing-SAME.
However, the feature-based ranking model, \texttt{Coor-Ascent}, performs better than all neural baselines on all three testing scenarios. The differences can be as high as $15\%$ and some are statistically significant.  This holds even for Testing-SAME which is expected to favor deep models that access more in-domain training data. These results remind that no `deep learning magic' can instantly provide significant gains for information retrieval tasks. The development of neural IR models also requires an understanding of the advantages of neural methods and how their advantages can be incorporated to meet the needs of information retrieval tasks.

\begin{figure*}[h]
\centering
\begin{subfigure}{0.29\textheight}
% {0.33\textwidth}
\includegraphics[width=\textwidth]{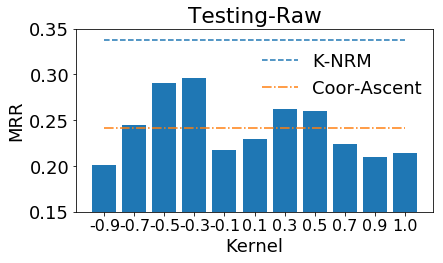}
\caption{Individual kernel's performance. \label{fig:kernel_performance}}
\end{subfigure}
\begin{subfigure}{0.29\textheight}
% {0.325\textwidth}
\includegraphics[width=\textwidth]{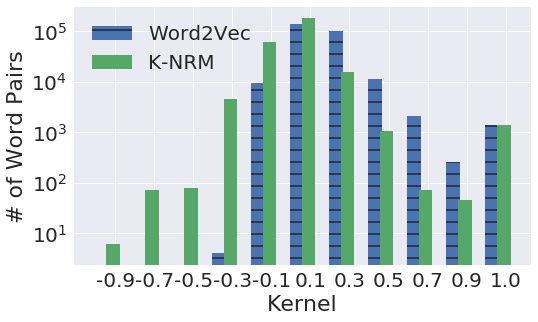}
\caption{Kernel sizes before and after learning. \label{fig:kernel_size}}
\end{subfigure}
\begin{subfigure}{0.21\textheight}
% {0.24\textwidth}
\includegraphics[width=\textwidth]{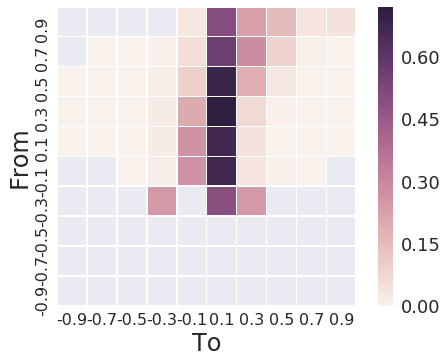}
\caption{Word pair movements. \label{fig:word_move}}
\end{subfigure}

\caption{Kernel guided word embedding learning in \texttt{K-NRM}.  
Fig.~\ref{fig:kernel_performance} shows the performance of \texttt{K-NRM} when only one kernel is used in \emph{testing}. Its X-axis is the $\mu$ of the used kernel. Its Y-axis is the MRR results.
Fig.~\ref{fig:kernel_size} shows the log number of word pairs that are closest to each kernel, before \texttt{K-NRM} learning (Word2Vec) and after. Its X-axis is the $\mu$ of kernels.
Fig.~\ref{fig:word_move} illustrates the word pairs' movements in \texttt{K-NRM}'s learning. The heat map shows the fraction of word pairs from the row kernel (before learning, $\mu$ marked on the left) to the column kernel (after learning, $\mu$  at the bottom).
\label{fig:kernel_learning}
}

\end{figure*}

\subsection{Source of Effectiveness}
\texttt{K-NRM} differs from previous ranking methods in several ways: multi-level soft matches, word embeddings learned directly from ranking labels, and the kernel-guided embedding learning. This experiment studies these effects by comparing the following variants of \texttt{K-NRM}.

\texttt{K-NRM (exact-match)} only uses the exact match kernel $(\mu, \sigma) = (1, 0.001)$. It is equivalent to TF.

\texttt{K-NRM (word2vec)} uses pre-trained word2vec, the same as \texttt{DRMM}. Word embedding is fixed; only the ranking part is learned. 

\texttt{K-NRM (click2vec)} also uses pre-trained word embedding. But its word embeddings are trained on (query word, clicked title word) pairs. 
The embeddings are trained using skip-gram model with the same settings used to train \texttt{word2vec}.
These embeddings are fixed during ranking.

\texttt{K-NRM (max-pool)} replaces kernel-pooling with max-pooling. Max-pooling finds the maximum similarities between document words and each query word;
it is commonly used by neural network architectures.  In our case, given the candidate documents' high quality, the maximum is almost always 1, thus it is similar to TF.

\texttt{K-NRM (mean-pool)} replaces kernel-pooling with mean-pooling.
It is similar to \texttt{Trans} except that the embedding is trained by learning-to-rank.

All other settings are kept the same as \texttt{K-NRM}.
Table~\ref{tab:variant} shows their evaluation results, together with the \texttt{full model} of \texttt{K-NRM}.

\emph{Soft match is essential.} \texttt{K-NRM (exact-match)} performs similarly to \texttt{Lm} and \texttt{BM25}, as does \texttt{K-NRM (max-pool)}. 
This is expected: without soft matches, the only signal for \texttt{K-NRM} to work with is effectively the TF score.

\emph{Ad-hoc ranking prefers relevance based word embedding.} Using
\texttt{click2vec} performs about $5$-$10\%$ better than using \texttt{word2vec}. User clicks are expected to be a better fit as they represent user search preferences, instead of word usage in documents. The relevance-based word embedding is essential for neural models to outperform feature-based ranking. 
\texttt{K-NRM (click2vec)} consistently outperforms \texttt{Coor-Ascent}, but \texttt{K-NRM (word2vec)} does not. 

\emph{Learning-to-rank trains better word embeddings}. \texttt{K-NRM} with \texttt{mean-pool} performs much better than \texttt{Trans}. 
They both use average embedding similarities; the difference is that \texttt{K-NRM (mean-pool)} uses the ranking labels to tune the word embeddings, while \texttt{Trans} keeps the embeddings fixed. The trained embeddings improve the ranking accuracy, especially on top ranking positions. 

\emph{Kernel-guided embedding learning provides better soft matches.} \texttt{K-NRM} stably outperforms all of its variants.
\texttt{K-NRM (click2vec)} uses the same ranking model, and its embeddings are trained on click contexts. \texttt{K-NRM (mean-pool)} also learns the word embeddings using learning-to-rank. The main difference is how the information from relevant labels is used when learning word embeddings. In \texttt{KNRM (click2vec)} and \texttt{K-NRM (mean-pool)}, training signals from relevance labels are propagated \emph{equally}  to all query-document word pairs.
In comparison, \texttt{K-NRM} uses kernels to enforce multi-level soft matches; query-document word pairs on different similarity levels are adjusted differently   (see Section~\ref{sec:model_optimization}). 

Table~\ref{tab:example_kernel} shows an example of \texttt{K-NRM}'s learned embeddings. The \textbf{bold} words in each row are those `activated' by the corresponding kernel: their embedding similarities to the query word `Maserati' fall closest to the kernel's $\mu$. The example illustrates that the kernels recover different levels of relevance matching: $\mu=1$ is exact match; $\mu=0.7$ matches the car model with the brand; $\mu=0.3$ is about the car color; $\mu=-0.1$ is background noise. The \texttt{mean-pool} and \texttt{click2vec}'s uni-level training loss mix the matches at multiple levels, while the kernels provide more fine-grained training for the embeddings.

\begin{table}[t]
\centering
\caption{Examples of word matches in different kernels. Words in {\bf{bold}} are those whose similarities with the query word fall into the corresponding kernel's range ($\mu$). \label{tab:example_kernel}}
\begin{tabular}{rl}
\hline
 $\mu$     &    Query: `Maserati'                                                               "    \\ \hline
 1.0  & {\bf{Maserati}}  {\color{mygray}{Ghibli black interior  \_ who knows}} \\
0.7  & {\color{mygray}{Maserati}} {\bf{Ghibli}} {\color{mygray}{black interior  \_ who knows}} \\
0.3 & {\color{mygray}{Maserati Ghibli}} {\bf{black}} interior   \_ {\bf{who}} {\color{mygray}{knows}} \\
-0.1 & {\color{mygray}{Maserati Ghibli black interior   \_ who}} {\bf{knows}} \\ 
%  $< -0.3$ & {\color{mygray}{Maserati Ghibli black interior   \_ who knows}} \\ 
 \hline
\end{tabular}
\end{table}

\subsection{Kernel-Guided Word Embedding learning}

In \texttt{K-NRM}, word embeddings are initialized by \texttt{word2vec} and trained by the kernels to provide effective soft-match patterns. 
This experiment studies
how training affects the word embeddings, showing the responses of kernels in ranking,
the word similarity distributions, and the word pair movements during learning.

Figure~\ref{fig:kernel_performance} shows the performance of \texttt{K-NRM} when only a single kernel is used
during \emph{testing}.
 The x-axis is the $\mu$ of the kernel. 
 The results indicate the kernels' importance.  The kernels on the far left ($\leq -0.7$), the far right ($\geq 0.7$), and in the middle ($\{-0.1, 0.1\})$ contribute little; the kernels on the middle left ($\{-0.3, -0.5\} $) contribute the most, followed by those on the middle right ($\{0.3, 0.5\}$). Higher $\mu$ does not necessarily mean higher importance or better soft matching. Each kernel focuses on a group of word pairs that fall into a certain similarity range; the importance of this similarity range is learned by the model.

Figure~\ref{fig:kernel_size} shows the number of word pairs activated in each kernel before training (\texttt{Word2Vec}) and after (\texttt{K-NRM}).
The X-axis is the kernel's $\mu$. The Y-axis is the log number of word pairs activated (whose similarities are closest to corresponding kernel's $\mu$). 
Most similarities fall into the range (-0.4, 0.6). These histograms help explain why the kernels on the far right and far left do not contribute much: because there are fewer word pairs in them.

Figure~\ref{fig:word_move} shows the word movements during the embedding learning. 
Each cell in the matrix contains the word pairs whose similarities are moved from the kernel in the corresponding row ($\mu$ on the left) to the kernel in the corresponding column ($\mu$ at the bottom). The color indicates the fraction of the moved word pairs in the original kernel. Darker indicates a higher fraction. 
Several examples of word movements are listed in Table~\ref{tab:moved_examples}. 
Combining Figure~\ref{fig:kernel_learning} and Table~\ref{tab:moved_examples}, the following trends can be observed in the kernel-guided embedding learning process.

\emph{Many word pairs are decoupled.} 
Most of  the word movements are from other kernels to the `white noise' kernels $\mu \in \{-0.1, 0.1\}$.  These word pairs are considered related by \texttt{word2vec} but not by \texttt{K-NRM}.
This is the most frequent effect in \texttt{K-NRM}'s embedding learning. Only about $10\%$ of word pairs with similarities $\geq 0.5$ are kept.
This implies that document ranking requires a stricter measure of soft match. For example, as shown in Table~\ref{tab:moved_examples}'s first row, a person searching for `China-Unicom', one of the major mobile carriers in China, is less likely interested in a document about `China-Mobile', another carrier; in the second row, `Maserati' and `car' are decoupled as  `car' appears in almost all candidate documents' titles, so it does not provide much evidence for ranking.

\emph{New soft match patterns are discovered.} \texttt{K-NRM} moved some word pairs from near zero similarities to important kernels.
As shown in the third and fourth rows of Table~\ref{tab:moved_examples}, there are word pairs that less frequently appear in the same surrounding context, but convey possible search tasks, for example, `the \emph{search} for \emph{MH370} '. \texttt{K-NRM} also discovers word pairs that convey strong `irrelevant' signals, for example, people searching for `BMW' are not interested in the `contact us' page.

% This depends on the weights in your learning to rank layer!!! This is implying that -0.3 has positive weight....
%
%
\emph{Different levels of soft matches are enforced.} Some word pairs moved from one important kernel to another. This may reflect the different levels of soft matches \texttt{K-NRM} learned.
Some examples are in the last two rows in Table~\ref{tab:moved_examples}.
The $-0.3$ kernel is the most important one, and received word pairs that encode search tasks; the $0.5$ kernel received synonyms, which are useful but not the most important, as exact match is not that important in our setting.

\begin{table}[t]
\centering
\caption{Examples of moved word pairs in \texttt{K-NRM}. \textbf{From} and \textbf{To} are the $\mu$ of the kernels the word pairs were in before learning (\texttt{word2vec}) and after (\texttt{K-NRM}). Values in parenthesis are the individual kernel's MRR on Testing-RAW, indicating the importance of the kernel. `$+$' and `$-$' mark the sign of the kernel weight $w_k$ in the ranking layer; `$+$' means word pair appearances in the corresponding kernel are positively correlated with relevance; `$-$' means negatively correlated.  }
\label{tab:moved_examples}
\begin{tabular}{r|r|l}
\hline
                                   \textbf{From} & \textbf{To} & \multicolumn{1}{c}{\textbf{Word Pairs}}                   \\ \hline
$\mu=0.9$             & $\mu=0.1$           & (wife, husband), (son, daughter),        \\
(0.20, $-$)     & (0.23, $-$) & (China-Unicom, China-Mobile)             \\ \hline
$\mu=0.5$             & $\mu=0.1$           & (Maserati, car),(first, time)            \\
(0.26, $-$)     & (0.23, $-$) & (website, homepage)                      \\ \hline
$\mu=0.1$              & $\mu=-0.3$           &  (MH370, search), (pdf, reader)   \\
(0.23, $-$)     & (0.30, $+$)   &  (192.168.0.1, router)    \\ \hline
$\mu=0.1$              & $\mu=0.3$           & (BMW, contact-us),      \\ 
(0.23, $-$)     & (0.26, $-$)   & (Win7, Ghost-XP) \\ \hline
$\mu=0.5$              & $\mu=-0.3$         &(MH370, truth), (cloud, share)                \\
(0.26, $-$)     & (0.30, $+$)   &   (HongKong, horse-racing)      \\ \hline
$\mu=-0.3$           & $\mu=0.5$            & (oppor9,  OPPOR),  (6080, 6080YY),       \\
(0.30, $+$)     & (0.26, $-$)   & (10086, www.10086.com)                   \\ \hline
\end{tabular}
\end{table}

\subsection{Required Training Data Size}

\begin{figure}[t]
\centering
\includegraphics[width=0.49\textwidth]{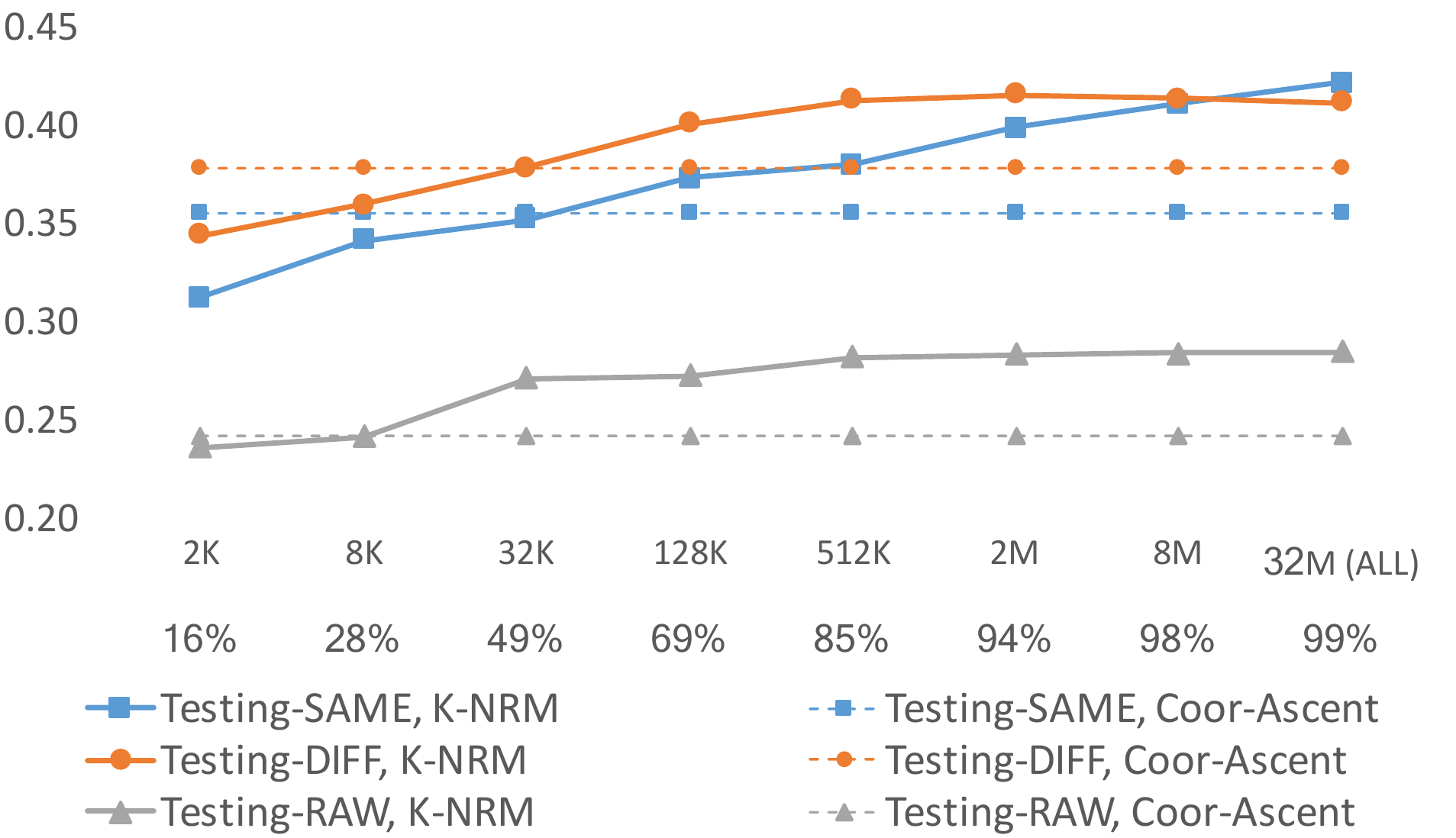}
\caption{\texttt{K-NRM}'s performances with different amounts of training data. X-axis: Number of sessions used for training, and the percentages of testing vocabulary covered (second row). Y-axis: NDCG@10 for Testing-SAME and Testing-DIFF, and MRR for Testing-RAW.}
\label{fig:train_size}
\end{figure}

This experiment studies \texttt{K-NRM}'s performance with varying amounts of training data.
Results are shown in Figure \ref{fig:train_size}.
The X-axis is the number of sessions used for training (e.g. $8K, 32K, \ldots$), and the coverage of testing vocabulary in the learned embedding (percentages).
Sessions were randomly sampled from the training set.  
The Y-axis is the performance of the corresponding model.
The straight and dotted lines are the performances of \texttt{Coor-Ascent}.

When only 2K training sessions are available, \texttt{K-NRM} performs worse than \texttt{Coor-Ascent}. Its word embeddings are mostly unchanged from \texttt{word2vec} as only 16\% of the testing vocabulary are covered by the training sessions. \texttt{K-NRM}'s accuracy grows rapidly with more training sessions. With only 32K ($0.1\%$) training sessions and $50\%$ coverage of the testing vocabulary, \texttt{K-NRM} surpasses \texttt{Coor-Ascent} on Testing-RAW. With 128K ($0.4\%$) training sessions and $69\%$ coverage on the testing vocabularies, \texttt{K-NRM} surpasses \texttt{Coor-Ascent} on Testing-SAME and Testing-DIFF. The increasing trends against Testing-SAME and Testing-RAW have not yet plateaued even with 31M training sessions, suggesting that \texttt{K-NRM} can utilize more training data. 
The performance on Testing-DIFF plateaus after 500K sessions, 
perhaps because the click models do not perfectly align with each other; 
more regularization of the \texttt{K-NRM} model might help under this
condition.

\subsection{Performance on Tail Queries}
This experiment studies how \texttt{K-NRM} performs on less frequent queries.  
We split the queries in the query log into Tail (less than 50 appearances), Torso (50-1000 appearances), and Head (more than 1000 appearances). For each category, 1000 queries are randomly sampled as testing; the remaining queries are used for training. Following the same experimental settings, the ranking accuracies of \texttt{K-NRM} and \texttt{Coor-Ascent} are evaluated. 

The results are shown in Table~\ref{tab:tail_queries}. Evaluation is only done using Testing-RAW as the tail queries do not provide enough clicks for DCTR and TACM to infer reliable relevance scores. The results show an expected decline of \texttt{K-NRM}'s performance on rarer queries. \texttt{K-NRM} uses word embeddings to encode the relevance signals, and as tail queries' words appear less frequently in the training data, it is hard to generalize the embedded relevance signals through them. 
Nevertheless, even on queries that appear less than $50$ times, \texttt{K-NRM} still outperforms \texttt{Coor-Ascent} by $8\%$.

\begin{table}[t]
\centering
\caption{Ranking accuracy on Tail (frequency $< 50$), Torso (frequency $50-1K$) and Head (frequency $> 1K$) queries. $\dagger$ indicates statistically significant improvements of \texttt{K-NRM} over \texttt{Coor-Ascent} on Testing-RAW. \textbf{Frac} is the fraction of the corresponding queries in the search traffic. \textbf{Cov} is the fraction of testing query words covered by the training data.
}
\label{tab:tail_queries}
\begin{tabular}{l|c|c|c|cc}
\hline
& \multirow{2}{*}{\textbf{Frac}}   
& \multirow{2}{*}{\textbf{Cov}}   
& \multicolumn{3}{c}{\textbf{Testing-RAW}, \textbf{MRR}}      \\ \cline{4-6} 

& & & \texttt{Coor-Ascent}   & \multicolumn{2}{c}{\texttt{K-NRM}} \\ \hline
Tail & 52\%  & $85\%$
& 0.2977    & 0.3230$^{\dagger}$      & +8.49\%      
\\

Torso & 20\%  & $91\%$
& 0.3202        & 0.3768$^{\dagger}$      & +17.68\%     
\\
Head & 28\%  & $99\%$
& 0.2415        & 0.3379$^{\dagger}$      & +39.92\%     \\ \hline
\end{tabular}
\end{table}

 \begin{figure}[t]
\centering
 \begin{subfigure}[!htb]{0.49\textwidth}
   \includegraphics[width=\textwidth]{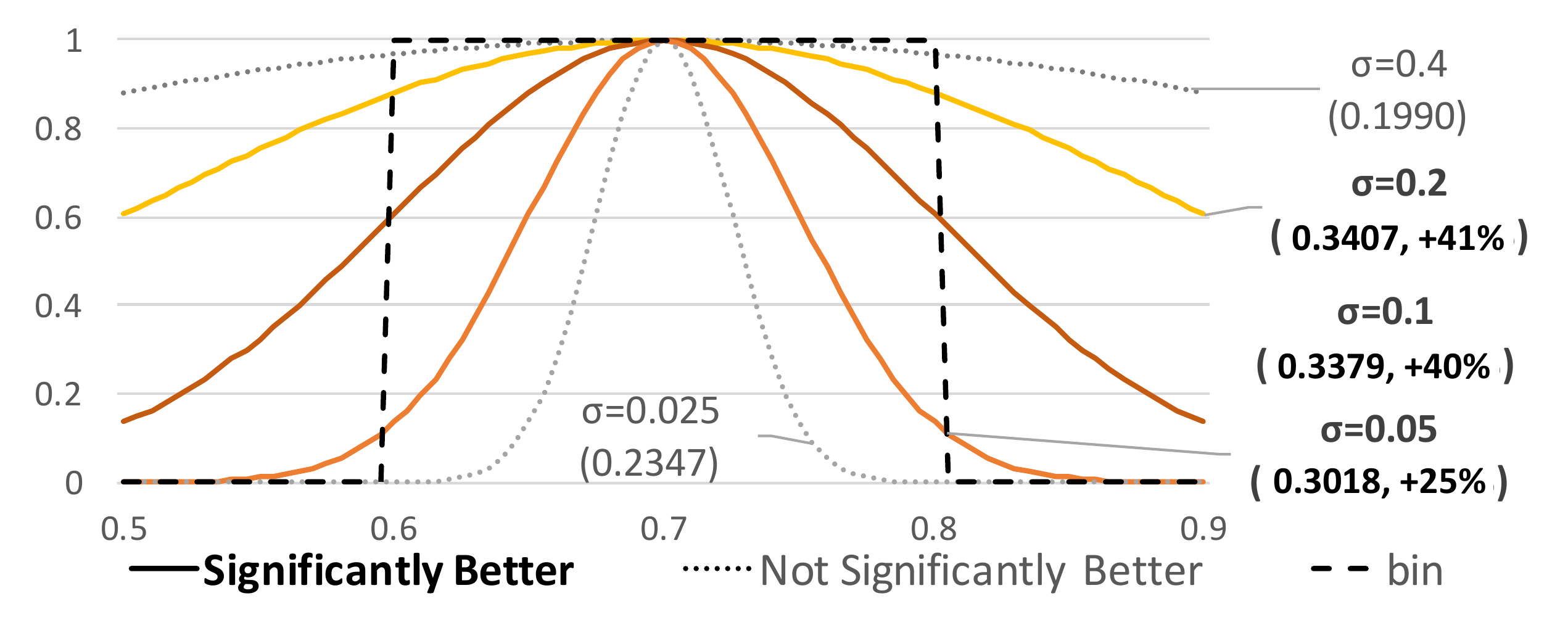}
\end{subfigure}
\caption{\texttt{K-NRM}'s performance with different $\sigma$.
MRR and relative gains over \texttt{Coor-Ascent} are shown in parenthesis.
Kernels drawn in solid lines indicate statistically significant improvements over \texttt{Coor-Ascent}. 
}
\label{fig:sigma}
\end{figure}

 \subsection{Hyper Parameter Study}
 \label{sec:sigma}
This experiment studies the influence of the kernel width ($\sigma$).
We varied the $\sigma$ used in \texttt{K-NRM}'s kernels, kept everything else unchanged, and evaluated its performance.
The shapes of the kernels with 5 different $\sigma$ and the corresponding ranking accuracies are shown in Figure \ref{fig:sigma}. Only Testing-RAW is shown due to limited space; the observation is the same on Testing-SAME and Testing-DIFF. 

As shown in Figure \ref{fig:sigma}, kernels too sharp or too flat either do not cover the  similarity space well, or mixed the matches at different levels; they cannot provide reliable improvements. With $\sigma$ between 0.05 and 0.2, $\texttt{K-NRM}$'s improvements are stable. 

We have also experimented with several other structures for \texttt{K-NRM}, for example, using more learning to rank layers, and using IDF to weight query words when combining their kernel-pooling results. However we have only observed similar or worse performances.  Thus, we chose to present the simplest successful model  to better illustrate the source of its effectiveness.

% \newpage
\section{Conclusion}
This paper presents \texttt{K-NRM}, a kernel based neural ranking model for ad-hoc search. 
The model captures word-level interactions using word embeddings, and ranks documents using a learning-to-rank layer. 
The center of the model is the new kernel-pooling technique. 
It uses kernels to softly count word matches at different similarity levels and provide soft-TF ranking features. The kernels are differentiable and support end-to-end learning. Supervised by ranking labels, the learning of word embeddings is guided by the kernels with the goal of providing soft-match signals that better separate relevant documents from irrelevant ones. The learned embeddings encode the relevance preferences and provide effective multi-level soft matches for ad-hoc ranking.

Our experiments on a commercial search engine's query log demonstrated \texttt{K-NRM}'s advantages. On three testing scenarios (in-domain, cross-domain, and raw user clicks), \texttt{K-NRM} outperforms both feature based ranking baselines and neural ranking baselines by as much as $65\%$, and is extremely effective at the top ranking positions. The improvements are also robust: Stable gains are observed on head and tail queries, with fewer training data,  a wide range of kernel widths, and a simple ranking layer.

Our analysis revealed that \texttt{K-NRM}'s advantage is not from `deep learning magic' but the long-desired soft match between query and documents, which is achieved by the kernel-guided embedding learning. Without it, \texttt{K-NRM}'s advantage quickly diminishes: its variants with only exact match, pre-trained word2vec, or uni-level embedding training all perform significantly worse, and sometimes fail to outperform the simple feature based baselines.

Further analysis of the learned embeddings illustrated how \texttt{K-NRM} tailors them for ad-hoc ranking: More than $90\%$ of word pairs that are mapped together in word2vec are decoupled, satisfying the stricter definition of soft match required in ad-hoc search. Word pairs that are less correlated in documents but convey frequent search tasks are discovered and mapped to certain similarity levels. The kernels also moved word pairs from one kernel to another based on their different roles in the learned soft match.

Our experiments and analysis not only demonstrated the effectiveness of \texttt{K-NRM}, but also provide useful intuitions about the advantages of neural methods  and how they can be tailored for IR tasks.
We hope our findings will be explored in many other IR tasks and will inspire more advances of neural IR research in the near future.

\section{Acknowledgments}
This research was supported by National Science Foundation (NSF) grant IIS-1422676, a Google Faculty Research Award, and a fellowship from the Allen Institute for Artificial Intelligence.
We thank Tie-Yan Liu for his insightful comments and Cheng Luo for helping us crawl the testing documents.
Any opinions, findings, and conclusions in this paper are the authors' and do not necessarily reflect those of the sponsors. 

\newpage

\bibliographystyle{abbrv}
\normalsize
\fontsize{7pt}{7pt}\selectfont
\bibliography{citation}
\end{document}